\documentclass[aps,prb,twocolumn,superscriptaddress,floatfix,longbibliography]{revtex4-2}
\usepackage{orcidlink}
\usepackage{amsmath,amssymb} % math symbols
\usepackage{physics}
\usepackage{bm} % bold math font
\usepackage [autostyle, english = american]{csquotes}
\MakeOuterQuote{"}
\usepackage[english]{babel}
\usepackage{enumitem}
\usepackage{graphicx} % for figures
\usepackage{comment} % allows block comments
\usepackage{textcomp} % This package is just to give the text quote '
\usepackage{enumitem}
\setlist{noitemsep,leftmargin=*,topsep=0pt,parsep=0pt}

\usepackage{xcolor} % \textcolor{red}{text} will be red for notes
\definecolor{lightgray}{gray}{0.6}
\definecolor{medgray}{gray}{0.4}

\usepackage{braket}

\usepackage{hyperref}
\hypersetup{
colorlinks=true,
urlcolor= blue,
citecolor=blue,
linkcolor= blue,
}
\newcommand{\mytitle}{Noisy Braiding of Majorana Modes: A Comparison of Nanowire Trijunction and Quantum-Dot-Assisted Architectures}

\begin{document}

\title{\mytitle}
\author{Dibyajyoti Sahu\,\orcidlink{0000-0003-4739-7538}}
\email[]{dibyajyoti20@iiserb.ac.in}
\affiliation{Department of Physics, Indian Institute of Science Education and Research, Bhopal, India}

\author{Suhas Gangadharaiah\,\orcidlink{0000-0001-7834-9438}}
\email[]{suhasg@iiserb.ac.in}
\affiliation{Department of Physics, Indian Institute of Science Education and Research, Bhopal, India}
\date{\today}

\begin{abstract}
Majorana zero modes have emerged as one of the most promising platforms for topological quantum computation, since their non-Abelian braiding statistics allow quantum information to be encoded nonlocally and manipulated through braiding operations that are, in principle, protected against local perturbations. In practice, however, a braid is only as robust as its physical implementation: finite-time operation, residual couplings, and environmental noise can all convert local excitations into logical errors during the exchange process. Here, we address this question through a microscopic comparison of two representative braiding architectures, a nanowire trijunction and a quantum-dot-assisted setup, simulating the full time-dependent Bogoliubov--de Gennes dynamics under both noiseless and noisy conditions. We %decompose the resulting gate error into diabatic, operational, and noise-induced contributions, and 
 show that the dot-assisted architecture consistently achieves a lower error over a shorter timescale than the trijunction, owing to its more localized exchange mechanism. This advantage persists in the presence of noise, and a spatially resolved analysis further reveals that, in the dot-assisted geometry, fast noise localized on the dot produces a smaller error than equivalent noise on the wires, whereas slow, quasi-static noise on the dot becomes the dominant limitation. Taken together, these findings link the different error contributions directly to device geometry, pointing to concrete design principles for noise-resilient Majorana-based quantum gates.
\end{abstract}

%%%%%%%%%%%%%%%%%%%%%%%%%%%%%%%%%%%%%%%%%%%%%%%%%%%%%%%%%%%%%%%%%%%%%%%%%%%%%%%%%%%%%%%%%%%%%%%%%%%%%%%%%%%%%%%%%%%%%%%%%%%%%%%%%%%%%%%%%%%%%%%%%%%%%%%%%%%%%%%%%%%%%%%%%%%%%%%%%%%%%%%%%%%%%
\maketitle
\section{\label{sec:Introduction}Introduction}
Majorana zero modes in topological superconductors have attracted sustained attention as a route toward fault-tolerant quantum computation~\cite{kitaev_unpaired_2001, nayak_non-abelian_2008, alicea_non-abelian_2011, sarma_majorana_2015, lutchyn_majorana_2018, aasen_milestones_2016, aasen_roadmap_2025}, since their non-Abelian braiding statistics allow quantum information to be processed through adiabatic exchanges within a degenerate ground-state manifold rather than by local control of microscopic degrees of freedom. The non-Abelian nature of Majorana zero modes has itself been the subject of extensive theoretical and experimental investigation~\cite{cheng_nonadiabatic_2011, karzig_shortcuts_2015, pedrocchi_monte_2015,rahmani_optimal_2017,sekania_braiding_2017,nag_diabatic_2019,knapp_nature_2016, beenakker_search_2020, chen_non-abelian_2022, liu_fusion_2023, xu_dynamics_2023, pan_braiding_2024, miles_braiding_2025,yu_nonadiabatic_2025,quade_exchangeless_2025,hodge_altermagnet-superconductor_2025,nitsch_adiabatic_2025, frey_majorana_2025, hodge_fusion_2026}, motivated by the prospect of directly verifying braiding statistics that have no counterpart among conventional quasiparticles. In the ideal limit, such braids implement unitary transformations that realize logical gates in a topologically protected manner~\cite{ivanov_non-abelian_2001, sau_controlling_2011, bauer_topologically_2019, plugge_majorana_2017}, and it is this feature that makes Majorana-based architectures especially appealing for robust quantum information processing. However, a central question is how far this ideal protection survives in realistic settings, where the braid is performed over finite time and in the presence of noisy controls~\cite{zhang_effects_2019, pedrocchi_majorana_2015, sahu_transport_2025}.

A variety of braiding architectures have been proposed for Majorana zero modes, including nanowire trijunction networks~\cite{alicea_non-abelian_2011,halperin_adiabatic_2012, torres_luna_design_2024, boross_braiding-based_2024}, Coulomb-assisted schemes~\cite{heck_coulomb-assisted_2012, plugge_majorana_2017}, and quantum-dot-based architectures~\cite{malciu_braiding_2018,sau_controlling_2011,liu_minimal_2021, tsintzis_majorana_2024}, each aiming to realize non-Abelian exchange through experimentally implementable control protocols. However, much of the existing literature is formulated within an effective low-energy Majorana description\cite{nag_diabatic_2019}, where the emphasis is placed on topological exchange operations, adiabatic evolution, and the resulting unitary transformations within the low energy manifold, rather than on the full microscopic dynamics of realistic devices. More recently, Mascot \emph{et al.} developed a many-body framework for simulating Majorana braiding and analyzed the fidelity of finite-time $X$- and $Z$-gate protocols in finite systems~\cite{mascot_many-body_2023}. On the noise side, Pedrocchi and DiVincenzo showed that thermal environments can convert local quasiparticle excitations into logical errors during braiding~\cite{pedrocchi_majorana_2015}, while Alase \emph{et al.} notably identified $1/f$ noise as an important source of decoherence in Majorana-based systems~\cite{alase_decoherence_2025}. Despite these advances, a systematic comparison of how different braiding geometries respond to fluctuating control parameters, particularly in the presence of $1/f$-type noise during finite-time braiding protocols, has received comparatively little attention.

In this work, we investigate the dynamics of finite-time Majorana braiding, including the quasiparticle excitations and other potential sources of error, through a microscopic comparison of two representative braiding geometries, a nanowire trijunction and a quantum-dot-assisted architecture. The error of the $X$ gate is quantified in both noiseless and noisy environments using time-dependent simulations of the full Bogoliubov--de Gennes Hamiltonian, in the presence of telegraph and $1/f$ noise. Interestingly, we find that while both geometries exhibit an optimal braiding timescale in the presence of noise, the dot-assisted protocol consistently achieves lower diabatic errors by operating over shorter timescales. Moreover, we show that the influence of noise depends strongly on its spatial location and temporal characteristics, leading to markedly different error responses in the two geometries. These results provide a microscopic understanding of the interplay between finite-time dynamics, device architecture, and environmental noise in Majorana braiding protocols.

The organization of the paper is as follows. In Sec.~\ref{sec:model}, we introduce the microscopic model and describe the qubit encoding used throughout this work. Section~\ref{sec:braiding_geometry} presents the two braiding geometries considered, namely the trijunction network and the quantum-dot-assisted protocol. In Sec.~\ref{sec:noise}, we introduce the noise models and investigate their impact on the braiding dynamics, separately examining the effects of global and spatially localized fluctuations. Finally, we summarize our main results and discuss their implications for Majorana-based topological quantum computation in Sec.~\ref{sec:conclusion}.
\section{Model\label{sec:model}}
We model Majorana zero modes (MZMs) using the one-dimensional spinless $p$-wave
Kitaev chain, which provides a minimal lattice description of a topological
superconductor~\cite{kitaev_unpaired_2001}. The Hamiltonian of an
$N$-site chain is
\begin{equation}
\begin{aligned}
H = - \sum_{j} & \left( t\, c_j^\dagger c_{j+1}
+ \Delta\, c_j c_{j+1} + \text{h.c.} \right) \\
& - \sum_{j} \mu_j \left( c_j^\dagger c_j - \frac{1}{2} \right),
\end{aligned}
\end{equation}
% \begin{equation}
% \begin{aligned}
% H = - \sum_{j} & \left( t\, c_j^\dagger c_{j+1}
% + \Delta\, c_j c_{j+1} + \text{h.c.} -\mu_j  c_j^\dagger c_j + \frac{\mu_j }{2} \right),
% \end{aligned}
% \end{equation}
where $t$ is the nearest-neighbor hopping amplitude, $\Delta$ is the induced
$p$-wave superconducting pairing, and $\mu_j$ denotes the (possibly
site-dependent) chemical potential. The fermionic operators are expressed in
terms of Majorana operators via $c_j = (\gamma_{2j-1} + i\gamma_{2j})/2$, with
$\gamma = \gamma^\dagger$, a decomposition that provides a transparent
description of the low-energy degrees of freedom. The model realizes a
topological superconducting phase for $|\mu| < 2t$, separated from a trivial
phase for $|\mu| > 2t$ by a bulk gap closing. Unless otherwise stated, we use
$t=\Delta=1$ as the unit of energy throughout the remainder of this work, and
we fix the chemical potential at $\mu=0.05$ in the topological phase and
$\mu=8$ in the trivial phase of the nanowire. In the topological phase, MZMs
appear localized at the ends of the chain, with an overlap that decays
exponentially with system size as $\sim e^{-L/\xi}$. For sufficiently long
chains, this overlap becomes negligible, and the ground-state manifold is
consequently nearly degenerate. This degeneracy is what enables the nonlocal
encoding of quantum information using spatially separated MZMs, and it forms
the basis for the qubit construction discussed below~\cite{alicea_non-abelian_2011, sarma_majorana_2015}. For the remaining part of the paper we are going to chose this parameter. $t = \Delta$ and we fix delta = 1 and mu = 0.05 for topological region and 8 for trivial region of the nanowire.

\subsection{Single-qubit encoding and Pauli $X$ gate}
A single logical qubit is encoded using four spatially separated MZMs
$\gamma_1, \gamma_2, \gamma_3,$ and $\gamma_4$~\cite{alicea_non-abelian_2011, plugge_majorana_2017}. The complex fermion operators
$f_{12} = (\gamma_1 + i\gamma_2)/2$ and $f_{34} = (\gamma_3 + i\gamma_4)/2$ are
defined, and once the total fermion parity is fixed, the four-dimensional
Hilbert space reduces to a two-dimensional logical subspace. The logical basis states are chosen as
\begin{equation}
\ket{0_\mathrm{L}} = \ket{0_{12},0_{34}}, \quad \text{and} \quad
\ket{1_\mathrm{L}} = \ket{1_{12},1_{34}},
\end{equation}
where both states belong to the same parity sector. 
The exchange of Majorana modes acts as a unitary transformation within
this degenerate ground-state manifold. The exchange of the two inner Majoranas
$\gamma_2$ and $\gamma_3$ is generated by the operator
$R_{23}^\dagger = (1 + \gamma_2 \gamma_3)/\sqrt{2}$, which maps a logical basis
state into a coherent superposition~\cite{ivanov_non-abelian_2001}. In particular the operation,
\begin{equation}
R_{23}^\dagger \ket{0_\mathrm{L}}
= \frac{1}{\sqrt{2}} \left( \ket{0_\mathrm{L}} - i\ket{1_\mathrm{L}} \right),
\end{equation}
reflects the non-Abelian nature of the braid. Performing the same exchange a second time completes the braiding cycle and yields
\begin{equation}
(R_{23}^\dagger)^2 \ket{0_\mathrm{L}} = \ket{1_\mathrm{L}},
\end{equation}
up to an overall phase fixed by the fermion parity sector. Thus, exchanging the inner Majoranas twice implements a Pauli $X$ operation on the logical qubit, $\ket{0_\mathrm{L}} \leftrightarrow \ket{1_\mathrm{L}}$.

Since the focus of this work is on the quasiparticle excitations generated during finite-time braiding, deviations from ideal adiabatic evolution are quantified using the overlap between the time-evolved state and the target ground state.
Starting from $\ket{0_\mathrm{L}}$ at $t=0$, we define the error associated with the Pauli $X$ gate as~\cite{knapp_nature_2016, karzig_optimal_2015, hodge_characterizing_2025}
\begin{equation}
\text{Error} = 1 - \left| \langle 1_\mathrm{L} \vert \psi(\tau) \rangle \right|^2 ,
\end{equation}
where $\ket{\psi(\tau)}$ denotes the many-body state obtained after completing the braiding protocol in a finite time $\tau$. 
In the adiabatic limit and perfect gate operation, $\text{Error} \to 0$, while finite values of Error indicate a deviation of the final state from the target logical state $\ket{1_\mathrm{L}}$.

\section{Braiding Geometry\label{sec:braiding_geometry}}

\begin{figure}[t]
    \centering
    \includegraphics[width=\linewidth]{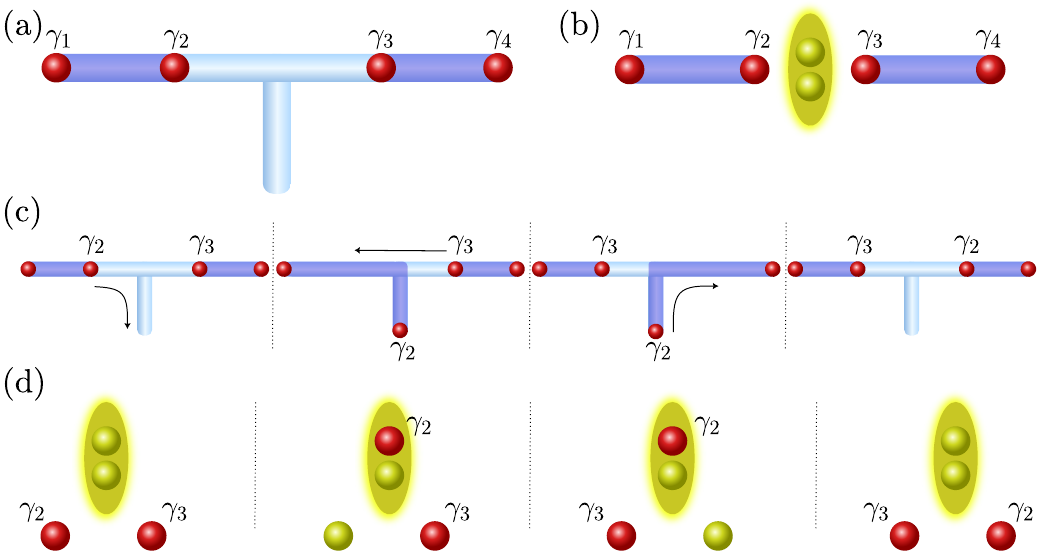}
    \caption{Braiding geometries and protocols. (a) Trijunction network and (b) quantum-dot-assisted architecture, showing the device layout to exchange the Majorana zero modes. (c),(d) Corresponding braiding steps for the trijunction and dot-assisted protocols, respectively, illustrating the sequence of modulations that implement the exchange.}
    \label{fig:braid_geom}
\end{figure}

To probe how finite-time dynamics affect the fidelity of Majorana braiding, we consider two representative device geometries that enable controlled exchange of MZMs.
The first geometry is a trijunction network formed by three Kitaev chains meeting at a central node~\cite{torres_luna_design_2024, alicea_non-abelian_2011, halperin_adiabatic_2012,boross_braiding-based_2024, pandey_majorana_2023}, which represents the minimal wire architecture allowing spatial braiding of MZMs. The second geometry consists of two topological nanowires coupled through a quantum dot~\cite{malciu_braiding_2018, sau_controlling_2011, liu_minimal_2021,zhang_quantifying_2025}, where the exchange of MZMs is mediated by tunable hybridization with the dot level.
These two setups realize braiding through fundamentally different mechanisms: in the trijunction geometry, MZMs are transported adiabatically along extended wire segments, whereas in the dot-assisted geometry the exchange occurs locally through controlled tunneling processes. A schematic illustration of both
architectures is shown in Fig.~\ref{fig:braid_geom}(a-b). %Comparing these two implementation schemes allows us to examine how the underlying braiding mechanism affects the susceptibility to diabatic excitations and noise-induced errors.
Comparing these two implementation schemes allows us to examine how the underlying braiding mechanism affects the susceptibility to diabatic excitations and noise-induced errors.

\subsection{Trijunction geometry}

We consider three one-dimensional Kitaev chains connected to a common central
site. Such a network allows the exchange of spatially separated MZMs by
circumventing the topological obstruction present in strictly one-dimensional
systems. Each leg of the trijunction is modeled as a spinless $p$-wave
superconducting chain with nearest-neighbor hopping amplitude $t$, pairing
amplitude $\Delta_p^{(n)} = |\Delta_p| e^{i\phi_n}$, and site-dependent
chemical potential $\mu_x$. The Hamiltonian of the full trijunction system is
\begin{equation}
\begin{aligned}
H = & - \sum_{x=0}^{3L} \mu_x c_x^\dagger c_x+\sum_{n=0}^{2} \Big[ -t\, c_0^\dagger c_{nL+1}
+ \Delta_p^{(n)} c_0^\dagger c_{nL+1}^\dagger + \mathrm{h.c.} \Big] \\
&+ \sum_{n=0}^{2} \sum_{x=nL+1}^{(n+1)L-1}
\Big[ -t\, c_x^\dagger c_{x+1}
+ \Delta_p^{(n)} c_x^\dagger c_{x+1}^\dagger + \mathrm{h.c.} \Big]
,
\end{aligned}
\end{equation}
where $n$ labels the three legs and $L$ denotes the length of each leg.

We set $L=20$ sites for each leg of the trijunction: two of the three legs
are prepared with their outermost 10 sites in the topological phase and the
inner 10 sites (nearest the junction) in the trivial phase, while the third
leg is held entirely in the trivial phase [see Fig.~\ref{fig:braid_geom}(a)].
Controlled transport of MZMs through the junction requires a suitable
superconducting phase difference between the three legs. We adopt the phase
convention $\phi_0 = 0$, $\phi_1 = \pi$, and $\phi_2 = \pi/2$, which
effectively realizes a $p_x + i p_y$ pairing structure at the junction \cite{alicea_non-abelian_2011, mascot_many-body_2023}. 
This choice avoids accidental zero-energy modes that may arise when multiple topological segments meet at a node\cite{pandey_majorana_2023}, and it ensures that the hybridization of MZMs at the junction remains well defined.
Such phase relations naturally emerge in nanowire networks with Rashba spin--orbit coupling, where wires oriented at right angles acquire a relative superconducting phase shift of $\pi/2$.

Braiding of MZMs along the trijunction is achieved by locally tuning the chemical potential using electrostatic gates. By driving a region of the wire between the topological ($|\mu|<2t$) and trivial ($|\mu|>2t$) regimes, the topological phase boundary, and hence the position of the localized MZM, can be displaced along the wire. %Braiding is achieved through a sequence of such local deformations that guide the MZMs through the junction.
To suppress diabatic excitations, the chemical potential is varied smoothly in time; the explicit ramping protocol is described in Appendix~\ref{app:mu_ramp}. Exchanging the two inner MZMs twice implements a topological Pauli $X$ gate on the encoded qubit [see Fig.\ref{fig:braid_geom}(c)]. We simulate this process by solving the time-dependent Bogoliubov--de Gennes equations and evaluating the overlap between the time-evolved state and the initial ground-state
manifold.

\begin{figure}[t]
    \centering
    \includegraphics[width=\linewidth]{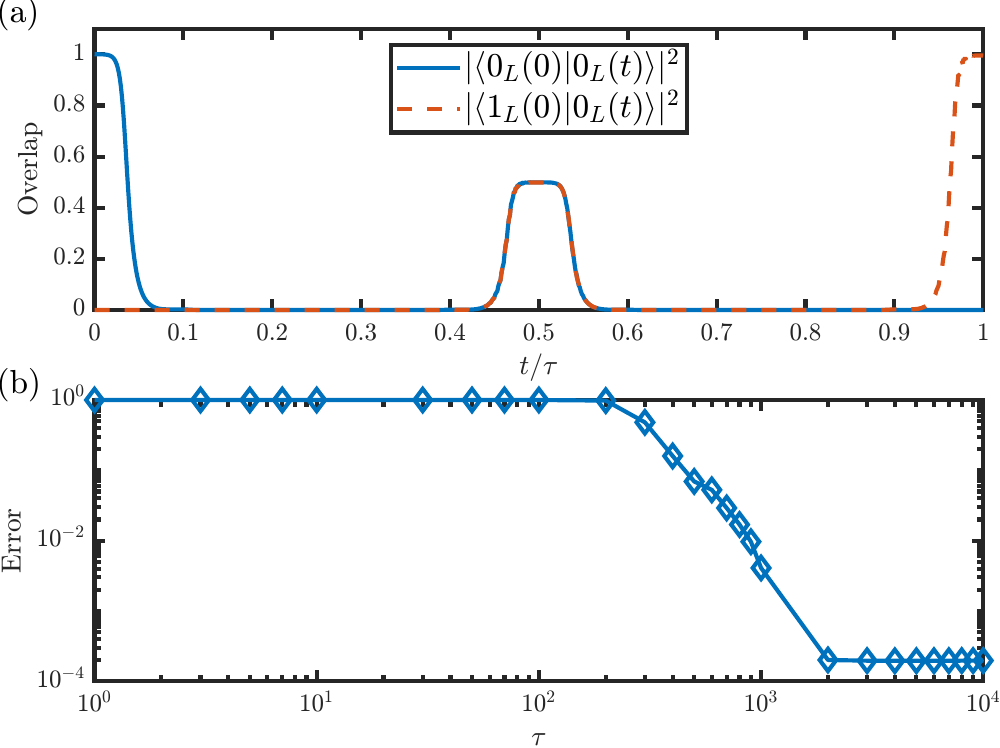}
    \caption{(a) Time evolution of the overlap during the trijunction braiding protocol implementing a Pauli $X$ operation. (b) Error as a function of the total braiding time, showing the suppression of the diabatic error and the saturation to a finite operational error at long drive times.}
    \label{fig:X_gate_overlap}
\end{figure}
Figure~\ref{fig:X_gate_overlap}(a) shows the time evolution of the overlap during the braiding protocol. It is clear from the figure that at the end of the adiabatic drive, the overlap between the time-evolved state $\ket{\psi(t)}$ and the target state $\ket{1_L}$ approaches unity, indicating that the MZMs are exchanged with negligible leakage from the ground-state manifold. This evolution realizes the desired Pauli $X$ operation on the encoded qubit.
The corresponding error is shown in Fig.~\ref{fig:X_gate_overlap}(b), which decreases systematically as the braiding duration increases. This trend reflects the suppression of the \emph{diabatic error} as the protocol becomes more adiabatic, reflecting the reduced nonadiabatic excitations generated as the drive is slowed. At long drive times, however, the total error saturates to a finite value instead of vanishing. This residual contribution is what we refer to as the \emph{operational error}: it arises not from the finite duration of the drive but from operating away from the Kitaev sweet spot, and it persists even in the fully adiabatic limit. %{\color{red}Precisely at the sweet spot, the operational error vanishes, and the total error continues to decrease with increasing drive time (not shown).}

\subsection{Dot-assisted braiding \label{sec:dot_noiseless}}
An alternative architecture for implementing Majorana braiding is provided by
two topological nanowires coupled through a quantum dot. In this geometry, the
dot replaces the central junction and acts as a tunable intermediate degree of
freedom that mediates the exchange of the end MZMs of the two wires. Such
dot-assisted setups provide a flexible route for manipulating MZMs using local
gate control while maintaining a simple device structure. % Each nanowire is modeled as a spinless $p$-wave superconducting chain with nearest-neighbor hopping amplitude $t$, pairing amplitude $\Delta_p^{(n)} = |\Delta_p| e^{i\phi_n}$, and chemical potential $\mu_x$, where $n=0,1$ labels the left and right wires.
The wires are connected through a quantum dot with a single electronic level
described by the fermionic operator $d$. The dot energy $\epsilon_d$ and the
tunnel couplings to the wire ends are taken to be tunable in time. The
effective Hamiltonian of the system can be written as
\begin{equation}
H = H_{\mathrm{w}} + H_{\mathrm{dot}} + H_{\mathrm{t}},
\end{equation}
where the first term represents the wire Hamiltonian, given by
\begin{equation}
\begin{aligned}
H_{\mathrm{w}} =
& \sum_{\substack{x=nL+1\\n=0,1}}^{(n+1)L-1}
\Big[ -t\, c_x^\dagger c_{x+1}
-\frac{\mu_x}{2} c_x^\dagger c_x+ \Delta_p^{(n)} c_x^\dagger c_{x+1}^\dagger + \text{h.c.} \Big] \\
%&- \sum_{x=1}^{2L} \mu_x c_x^\dagger c_x ,
\end{aligned}
\end{equation}
the dot Hamiltonian is given by
\begin{equation}
H_{\mathrm{dot}} = \epsilon_d d^\dagger d ,
\end{equation}
while the tunneling Hamiltonian between the dot and the wire ends is given by
\begin{equation}
H_{\mathrm{t}}
= \Big[t_L d^\dagger c_{1} + t_R d^\dagger c_{L+1} + \text{h.c.}\Big].
\end{equation}
Here, each nanowire is modeled as a spinless $p$-wave superconducting chain
with nearest-neighbor hopping amplitude $t$, pairing amplitude
$\Delta_p^{(n)} = |\Delta_p| e^{i\phi_n}$, and chemical potential $\mu_x$,
where $n=0,1$ labels the left and right wires, each of length $L=10$ sites.
%The wires are connected through a quantum dot with a single electronic level described by the fermionic operator $d$. The dot energy $\epsilon_d$ and the tunnel couplings to the wire ends are assumed to be tunable in time.
To ensure controlled hybridization of the MZMs, we choose the superconducting phases $\phi_0=0$ and $\phi_1=\pi$, which ensures that the dot couples to distinct Majorana components of the two wires~\cite{liu_minimal_2021, zhang_quantifying_2025}. 
With this phase choice, the dot
acts as a coherent bridge through which MZMs can be exchanged without closing
the bulk gap.

Braiding is implemented dynamically by modulating the dot level
$\epsilon_d(t)$ together with the tunnel couplings $t_L(t)$ and $t_R(t)$~\cite{liu_minimal_2021,zhang_quantifying_2025,miles_braiding_2025}.
Initially, the dot is far detuned ($\epsilon_d=E_0$) and decoupled from both the wires ($t_L=t_R=0$), leaving the MZMs localized at the wire ends. The exchange
protocol consists of three stages of duration $T$ each. First, the dot level
is brought closer to resonance while the coupling to the left wire is turned
on. Next, the MZM is transferred from the left wire to the right wire by
smoothly switching off $t_L$ and turning on $t_R$. Finally, the dot is
decoupled and its level restored to the initial value. This sequence completes
a non-Abelian exchange mediated by the quantum dot while preserving a finite
bulk gap throughout the evolution. Repeating the protocol twice implements the
Pauli $X$ gate in the adiabatic limit [see Fig.~\ref{fig:braid_geom}(d)].

\begin{figure}[t]
    \centering
    \includegraphics[width=\linewidth]{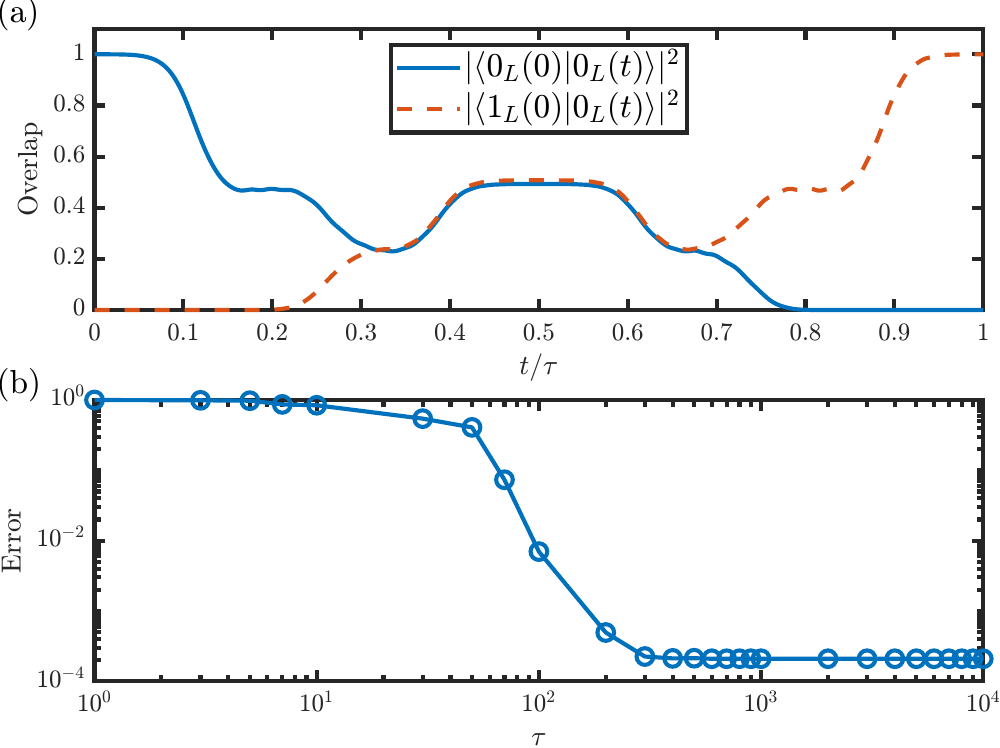}
   \caption{(a) Time evolution of the overlap during the quantum-dot-assisted
        braiding protocol mediated by a quantum dot. The overlap approaches unity
        at the end of the drive, indicating successful exchange within the
        ground-state manifold. (b) Error as a function of the total protocol
        duration, showing systematic suppression of the diabatic error as the
        drive time is increased, with saturation to a finite operational error
        at long drive times.}
    \label{fig:Dot_overlap}
\end{figure}

Figure~\ref{fig:Dot_overlap}(a) shows the overlap during the dot-assisted
braiding protocol. As in the trijunction case, the overlap approaches unity at
the end of the adiabatic drive, indicating that the exchange occurs with
minimal leakage from the ground-state manifold.
The error, shown in Fig.~\ref{fig:Dot_overlap}(b), decreases systematically with increasing drive time.  This trend reflects the suppression of the diabatic error as the protocol becomes more adiabatic, while the operational error remains finite at long times, as in the trijunction case. % and sets the asymptotic value the total error approaches.
Notably, a direct comparison with the trijunction setup reveals that the diabatic error in the dot-assisted protocol decreases more rapidly with drive time, reaching a comparably low value over a much shorter timescale. This improvement arises because the trijunction protocol requires adiabatic motion of MZMs along extended wire segments, whereas the dot-assisted scheme realizes the exchange through local hybridization, which is inherently less sensitive to diabatic excitations and therefore suppresses them more efficiently as the drive slows.

\section{Effect of Noise\label{sec:noise}}

This section investigates the effect of noise on the quasiparticle excitations generated during Majorana braiding. In realistic devices, temporal fluctuations in electrostatic gate potentials and the surrounding environment perturb the control parameters, introducing additional nonadiabatic effects that degrade braiding fidelity. To quantify these effects, we incorporate noise directly into the chemical potential within the braiding protocols described above. We consider two representative noise models: telegraph noise, modeled as a two-level fluctuator capturing the influence of charge traps, and $1/f$ noise, constructed as a superposition of telegraph processes with a broad distribution of switching rates. We examine both global noise, acting uniformly across the system, and spatially localized noise restricted to specific regions, in order to identify the parts of the device most susceptible to noise-induced quasiparticle excitations.

\subsection{Noise Models}

We model temporal fluctuations using dichotomous (telegraph) noise $\eta(t)$,
a two-state stochastic process that switches between values $a$ and $b$ with
rates $k_a$ and $k_b$, capturing charge noise and fluctuating impurities in
solid-state devices~\cite{abel_decoherence_2008,daniotti_qubit_2018,wold_decoherence_2012,krzywda_adiabatic_2020}.
In the stationary limit, the occupation probabilities are
$P_s(a)=k_b/(k_a+k_b)$ and $P_s(b)=k_a/(k_a+k_b)$, yielding a zero-mean
process with autocorrelation
$\langle\eta(t)\eta(t')\rangle=\sigma^2 e^{-|t-t'|/\tau_c}$, where
$\tau_c=1/(k_a+k_b)$ and
$\sigma^2=\frac{(a-b)^2k_ak_b}{(k_a+k_b)^2}$.
For simulations, we use symmetric telegraph noise with
$a=-\delta$, $b=\delta$, and $k_a=k_b=k_0$, giving $\tau_c=1/(2k_0)$.
Discrete noise trajectories are generated as described in
\cite{sahu_transport_2025}; for more details, see Appendix~B of
the same reference..

The Lorentzian spectrum of telegraph noise provides a microscopic building
block for low-frequency noise. In many solid-state devices, $1/f$ noise
arises from an ensemble of independent two-level fluctuators with a broad
distribution of switching rates, and we adopt this picture to construct
$1/f$ noise from telegraph processes. Each fluctuator with switching rate
$\gamma = 1/\tau_c$ produces a power spectral density
\begin{equation}
S_{\mathrm{TLF}}(f;\gamma)
= \frac{2\sigma_0^2 \gamma}{\gamma^2+f^2},
\end{equation}
where $\sigma_0^2$ is the variance of a single fluctuator. We assume a
distribution of switching rates
\begin{equation}
\mathcal{D}(\gamma) = \frac{N_1}{\gamma},
\qquad
\gamma \in [\gamma_{\min},\gamma_{\max}],
\end{equation}
where $N_1$ sets the overall noise strength. The total spectrum is then
\begin{equation}
\begin{aligned}
S_{1}(f)
&= \int_{\gamma_{\min}}^{\gamma_{\max}}
\mathcal{D}(\gamma)\,S_{\mathrm{TLF}}(f;\gamma)\,d\gamma . \\
\end{aligned}
\end{equation}
For frequencies in the intermediate regime
$\gamma_{\min} \ll f \ll \gamma_{\max}$, %this integral evaluates to
%\begin{equation}
%\begin{aligned}
%\int_{\gamma_{\min}}^{\gamma_{\max}}
%\frac{d\gamma}{\gamma^2+f^2}
%&= \frac{1}{f}
%\left[
%\arctan\!\left(\frac{\gamma_{\max}}{f}\right)
%- \arctan\!\left(\frac{\gamma_{\min}}{f}\right) \right]\\
%&\approx \frac{\pi}{2f}.
%\end{aligned}
%\end{equation}
the integral yields
\begin{equation}
S_{1}(f) \approx \frac{\pi \sigma_0^2 N_1}{f},
\end{equation}
which exhibits the characteristic $1/f$ dependence. %This construction shows how $1/f$ noise emerges from an ensemble of telegraph fluctuators with widely distributed switching times.
In what follows, we incorporate both the telegraph noise and the resulting
$1/f$ noise into the chemical potential to study their impact on
quasiparticle excitations during Majorana braiding.
\begin{figure}[t]
    \centering
    \includegraphics[width=\linewidth]{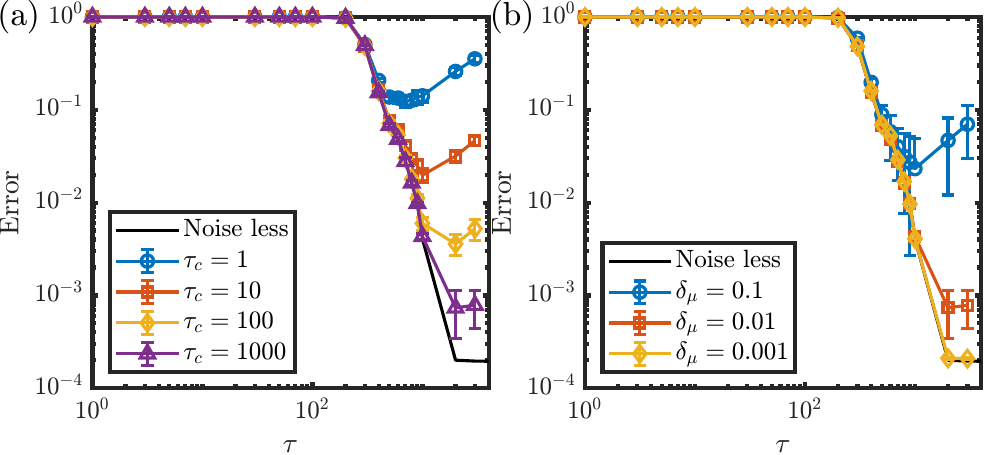}
   \caption{Error of the Pauli $X$ gate for the trijunction braiding
    protocol under global telegraph noise, averaged over multiple noise
    realizations, as a function of the total drive time. Panels (a) and (b)
    show the dependence on the noise correlation time and noise strength,
    respectively.}
    \label{fig:X_gate_noise_telegraph}
\end{figure}
\subsection{Global noise response}
We first analyze the effect of global noise, where temporal fluctuations are applied uniformly to the chemical potential on all sites of the device, providing a natural benchmark for comparing different braiding architectures. Figure~\ref{fig:X_gate_noise_telegraph} shows the total error for the trijunction braiding protocol under global telegraph noise, averaged over many independent noise realizations.
In contrast to the noiseless case, the total error no longer decreases monotonically with increasing drive time but instead develops a minimum at an optimal protocol duration.
This behavior reflects the competition between two components of the error: the diabatic error, which dominates at short drive times, and the noise-induced error, which accumulates and comes to dominate over longer evolution. %The position of this optimum depends on the noise parameters.
The position of the optimum drive time depends on the noise parameters: increasing either the noise correlation time or the noise strength shifts the minimum toward shorter drive times and raises the minimum achievable error.
Beyond a threshold noise strength [see Fig.~\ref{fig:X_gate_noise_telegraph}(b)], the error remains large over the entire range of drive times, indicating that high-fidelity braiding in the trijunction geometry is no longer attainable.

\begin{figure}[t]
    \centering
    \includegraphics[width=\linewidth]{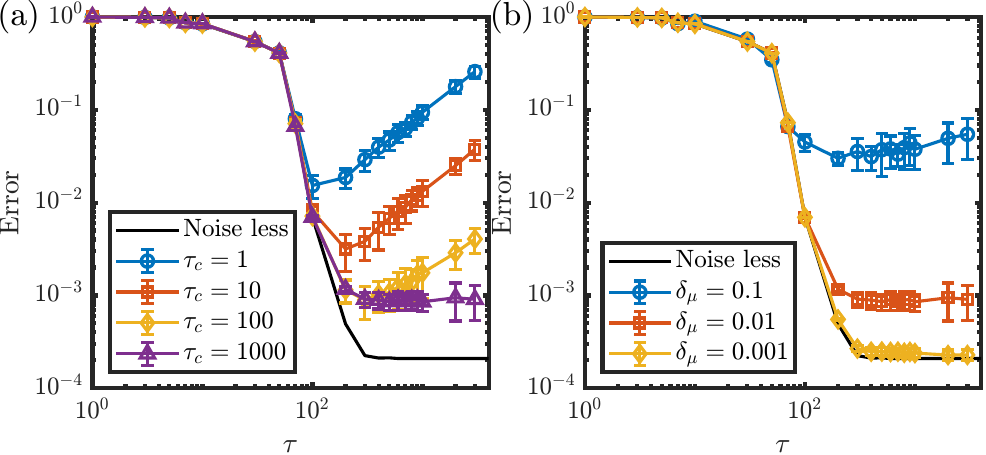}
    \caption{Error of the Pauli $X$ gate for the dot-assisted braiding
    protocol under global telegraph noise, averaged over multiple noise
    realizations, as a function of the total drive time. Panels (a) and (b)
    show the dependence on the noise correlation time and noise strength,
    respectively.}
    \label{fig:Dot_X_gate_noise_telegraph}
\end{figure}

Figure~\ref{fig:Dot_X_gate_noise_telegraph} shows the corresponding results for the dot-assisted braiding protocol. The qualitative behavior is similar, including the emergence of an optimal drive time. However, the minimum error is reached at shorter drive times compared to the trijunction setup. %This difference reflects the distinct braiding mechanisms.
This difference arises from the distinct braiding mechanisms, as discussed in Sec.~\ref{sec:dot_noiseless} for the noiseless case: since the dot-assisted scheme realizes the exchange through local hybridization rather than the extended adiabatic motion required in the trijunction, its diabatic-error component falls off more rapidly with drive time, allowing a comparably low total error to be reached over a much shorter protocol duration before the noise-induced component takes over.

\begin{figure}[t]
    \centering
    \includegraphics[width=\linewidth]{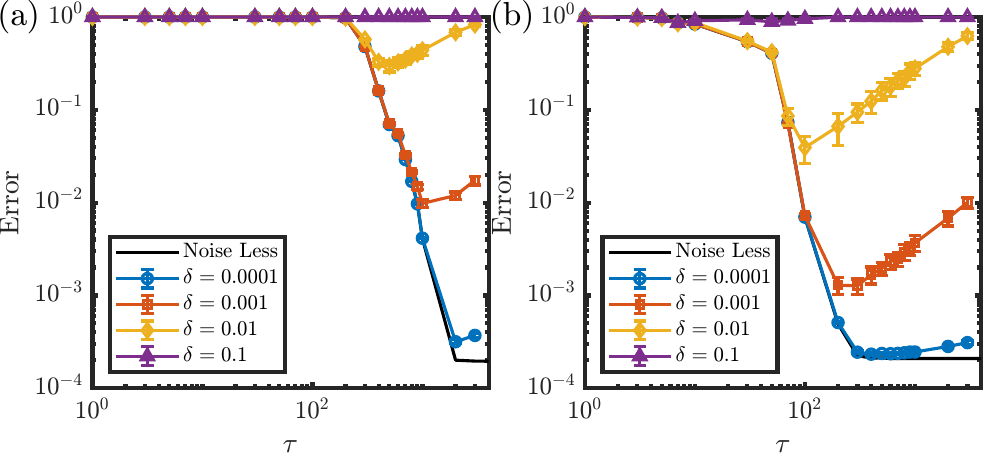}
    \caption{Error of the Pauli $X$ gate under global $1/f$ noise as a
    function of the total drive time, for different noise strengths.
    (a) Trijunction braiding protocol. (b) Dot-assisted braiding protocol.
    In both cases, results are averaged over multiple noise realizations.}
    \label{fig:1byf_noise_combined}
\end{figure}

We next consider global $1/f$ noise. Figure~\ref{fig:1byf_noise_combined}(a) shows the resulting total error for the trijunction geometry. Similar to the telegraph-noise case, the error exhibits a nonmonotonic dependence on the drive time, with a well-defined minimum set by the competition between the diabatic and noise-induced error components.
Notably, increasing the noise strength shifts this minimum toward shorter drive times and raises the minimum achievable error. Figure~\ref{fig:1byf_noise_combined}(b) shows the corresponding results for the dot-assisted geometry, where the qualitative behavior remains the same, with an optimal drive time that shifts systematically with noise strength.
However, the minimum error is reached at significantly shorter drive times compared to the trijunction setup. Thus, although both telegraph and $1/f$ noise impose similar constraints on braiding fidelity, the dot-assisted protocol retains a systematic advantage by operating at faster timescales, thereby reducing its exposure to the low-frequency fluctuations that dominate realistic noise environments.

\subsection{Local noise response}

\begin{figure*}
    \centering
    \includegraphics[width=\linewidth]{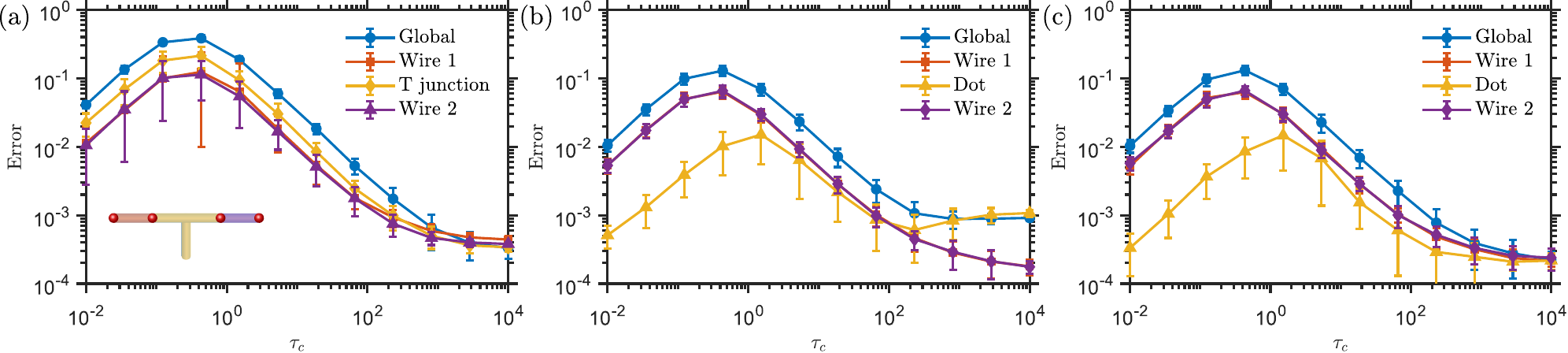}
   \caption{Error as a function of noise correlation time $\tau_c$ for a fixed
    drive rate, with noise applied selectively to the outer wire segments,
    the central region (junction or quantum dot), and globally.
    (a) and (b) show results for the trijunction and dot-assisted geometries,
    respectively, under symmetric telegraph noise switching between two
    finite values, $-\delta$ and $+\delta$. The inset in (a) shows a
    schematic of the trijunction device, with the outer wire segments and
    the central junction color-coded to match the corresponding error
    curves. (c) shows the dot-assisted geometry under telegraph noise
    switching between $0$ and $2\delta$, where one of the two states
    corresponds to zero noise.}

    \label{fig:tau_variation}
\end{figure*}

To identify which regions of the device are most susceptible to noise-induced quasiparticle excitations during braiding, we examine the effect of spatially localized noise. In contrast to the global-noise case, temporal fluctuations are restricted here to selected parts of the system, allowing us to isolate the dominant sources of diabatic error. In the trijunction geometry, the system consists of two outer wire segments
hosting the MZMs at the initial and final stages of the protocol, and a
central junction that forms the actively driven region where the control
parameters are varied in time [see inset of Fig.~\ref{fig:tau_variation}(a)]. For the dot-assisted case, the noise regions are
similarly divided into outer nanowire segments and the quantum dot region,
which serves as the central element mediating the exchange.

We fix a representative drive rate, chosen such that the diabatic error
contributes negligibly, and analyze the error as a function of the noise
correlation time $\tau_c$, with noise applied separately to the outer wire
segments, the central region, and all regions. The results are shown in
Figs.~\ref{fig:tau_variation}(a) and~\ref{fig:tau_variation}(b) for the
trijunction and dot-assisted setups, respectively. In the trijunction
geometry, noise acting on the junction consistently produces a larger error
than noise on the outer wire segments across all correlation times,
indicating that the noise-induced error is predominantly generated in the
actively driven region. In contrast, the dot-assisted geometry exhibits a
qualitatively different response. Over a broad range of correlation times,
noise acting on the quantum dot results in a smaller error than noise on the
wires, and only in the slow-noise limit does the dot become the dominant
source of error. This behavior reflects the hybridization mechanism
underlying the protocol: fast fluctuations average out over the duration of
the drive and have a reduced effect on the dot, whereas slow noise acts as a
quasi-static shift of the dot potential, moving it away from the optimal
operating point and disrupting the exchange process. Effectively, it is the
operational error that rises under slow noise, since the quasi-static shift
displaces the dot potential from the sweet spot. To verify this interpretation, we
replace the symmetric fluctuations in the range $\pm\delta$ with an
asymmetric interval $[0,2\delta]$ and repeat the analysis. In this case, the
enhancement of the error in the slow-noise regime is no longer observed [see
Fig.~\ref{fig:tau_variation}(c)], confirming that the effect originates from
the quasi-static nature of the noise and its interplay with the
hybridization dynamics. Overall, the impact of noise is governed by both its
spatial location and its temporal structure. In the trijunction geometry,
the junction remains the dominant source of error across all regimes. In
contrast, the dot-assisted protocol exhibits a crossover behavior in which
fast fluctuations on the dot are effectively suppressed, while slow,
quasi-static give rise to operational error and degrade braiding fidelity.
% This suggests that noise mitigation strategies should be tailored to the specific architecture: for trijunction devices, suppressing fluctuations in the junction is essential across all timescales, whereas in dot-assisted setups, low-frequency noise on the quantum dot is particularly detrimental, while high-frequency fluctuations are comparatively less harmful.

\section{Discussion \& Conclusion\label{sec:conclusion}}
In this work, we investigated the dynamics of finite-time Majorana braiding, including the quasiparticle excitations and other sources of error, by analyzing two representative geometries: a trijunction network and a quantum-dot-assisted nanowire setup. Both architectures allow non-Abelian exchange within the Kitaev-chain framework, but differ in the physical mechanism underlying the braid. The trijunction protocol relies on the adiabatic motion of MZMs along extended wire segments, whereas the dot-assisted scheme implements the exchange through local, tunable hybridization. This distinction between an extended and a local exchange mechanism is precisely what makes the two geometries a useful comparison for examining how diabatic effects and noise couple to the braiding dynamics.

In the absence of noise, both geometries approach the ideal adiabatic limit as the protocol duration is increased, exhibiting a systematic suppression of the diabatic error. However, the dot-assisted protocol consistently reaches a comparably low error over a much shorter timescale. This is because the trijunction relies on extended avoided crossings along the wire, while the dot-assisted scheme is governed by localized hybridization with a larger effective gap, leading to reduced nonadiabatic excitations. In both geometries, the total error saturates to a finite value at long drive times; this residual operational error reflects deviation from the Kitaev sweet spot rather than the finite duration of the drive, and vanishes only when the sweet spot is reached exactly.

In the presence of global noise, the total error acquires a nonmonotonic dependence on the drive time, with an optimal protocol duration set by the competition between the diabatic and noise-induced error components. This behavior is observed for both telegraph and $1/f$ noise. Despite this added complexity, the dot-assisted geometry retains its advantage: because it reaches its optimal operating point at shorter times, it is exposed to slow fluctuations for a correspondingly shorter duration.

A more detailed picture emerges when considering spatially localized noise. For the trijunction geometry, noise acting on the junction dominates the error across all correlation times, indicating that the noise-induced error is primarily generated in the actively driven region. In contrast, the dot-assisted setup exhibits a crossover behavior governed by the noise correlation time. Fast fluctuations on the dot are effectively averaged out during the protocol and lead to reduced error, whereas slow, quasi-static noise shifts the dot potential away from its optimal value, effectively raising the operational error and significantly degrading the braiding fidelity. This sensitivity is directly linked to the hybridization mechanism that mediates the exchange, and it depends on the temporal structure of the noise.

These results demonstrate that the impact of noise on Majorana braiding is controlled not only by its strength and spatial location but also by its temporal structure and its coupling to the underlying control mechanism. While both geometries are affected by similar sources of noise, the dot-assisted protocol provides a systematic advantage by enabling faster operations and reducing sensitivity to high-frequency fluctuations. At the same time, its performance is strongly influenced by low-frequency noise on the quantum dot, highlighting the importance of targeted noise mitigation in the actively controlled region.
Overall, our analysis identifies key mechanisms governing diabatic, operational, and noise-induced errors in Majorana braiding protocols and provides guidelines for optimizing their performance in realistic devices. In particular, minimizing low-frequency noise in the control elements and designing protocols that operate at shorter timescales can significantly enhance braiding fidelity. These considerations are expected to be important for the implementation of scalable topological quantum gates in solid-state platforms.

\begin{acknowledgments}
    We would like to thank Themba Hodge and Prof. Stephan Rachel for their help in clarifying several questions during the early stages of this project.
\end{acknowledgments}
\appendix
\section{Chemical-potential modulation protocol}
\label{app:mu_ramp}

In this appendix, we provide the explicit protocol used to transport Majorana zero modes by dynamically tuning the local chemical potential. The approach follows established "keyboard" or "piano-key" schemes, in which spatially resolved gate voltages allow individual sites of a Kitaev chain to be driven between the topological and trivial regimes.

The trijunction is operated by varying the on-site chemical potential $\mu_x(t)$ such that the topological regions of the wires expand or contract in time. A Majorana zero mode remains localized at the boundary between topological ($|\mu|<2t$) and trivial ($|\mu|>2t$) segments, and its position follows the motion of this boundary.

To minimize diabatic transitions into bulk excited states, the chemical potential is ramped smoothly using a smoothstep function
\begin{equation}
r(q)=
\begin{cases}
0, & q \le 0,\\
q^2(3-2q), & 0 \le q \le 1,\\
1, & q \ge 1,
\end{cases}
\end{equation}
which ensures that the first time derivative vanishes at the beginning and end of the ramp.

For a leg consisting of $L$ sites, indexed by $x=0,\ldots,L-1$, the chemical potential at site $x$ is taken to be
\begin{equation}
\mu_x(t) = \mu_{\mathrm{triv}}
+ (\mu_{\mathrm{topo}}-\mu_{\mathrm{triv}})
\, r\!\left(
\frac{t}{\tau(1+\alpha(L-1))}-\alpha x
\right),
\end{equation}
where $\tau$ sets the characteristic ramp duration and $\alpha$ is a delay coefficient that controls the relative timing between neighboring sites. The limit $\alpha=0$ corresponds to simultaneous modulation of all sites on a given leg, while $\alpha\gtrsim 1$ implements a strictly sequential, site-by-site ramping protocol. The reverse process is obtained by replacing $r \rightarrow 1-r$.

This parametrization allows us to interpolate continuously between fully collective and fully local transport of Majorana modes, and it provides a convenient control knob for studying the trade-off between adiabaticity and protocol duration.

\section{Bloch–Messiah decomposition}
\label{app:bloch_messiah}

In this appendix, we present the formalism used to compute many-body overlaps between the ground and excited states of quadratic fermionic Hamiltonians~\cite{mascot_many-body_2023}. This framework plays a central role in our analysis of dynamical fidelity, diabatic errors, and gate operations involving Majorana bound states. The discussion here closely follows standard treatments of Bogoliubov--de Gennes (BdG) systems, while emphasizing aspects relevant for numerical implementation and time-dependent problems.

We begin by considering a generic quadratic Hamiltonian written in the Nambu (BdG) basis,
\begin{equation}
	\mathcal{H}
	=
	\frac{1}{2}
	\begin{pmatrix}
		c^\dagger & c
	\end{pmatrix}
	\begin{pmatrix}
		H & \Delta \\
		\Delta^\dagger & -H^*
	\end{pmatrix}
	\begin{pmatrix}
		c \\
		c^\dagger
	\end{pmatrix},
\end{equation}
where $c^\dagger$ and $c$ denote vectors of fermionic creation and annihilation operators, respectively. The matrix structure ensures particle--hole symmetry and allows the Hamiltonian to be diagonalized by a Bogoliubov transformation. Introducing quasiparticle operators $d_k$, we relate the fermionic operators to the quasiparticles via
\begin{equation}
	\begin{pmatrix}
		c \\
		c^\dagger
	\end{pmatrix}
	=
	\begin{pmatrix}
		U & V^* \\
		V & U^*
	\end{pmatrix}
	\begin{pmatrix}
		d \\
		d^\dagger
	\end{pmatrix},
\end{equation}
where $U$ and $V$ satisfy the usual Bogoliubov unitarity constraints. The ground state $\ket{\Psi_d}$ is defined as the vacuum of all quasiparticle annihilation operators,
\begin{equation}
	d_k\ket{\Psi_d} = 0 \qquad \forall k .
\end{equation}

A formal expression for the ground state in terms of the fermionic vacuum $\ket{0_c}$ can be written as
\begin{equation}
	\ket{\Psi_d}
	=
	\frac{1}{\sqrt{\mathcal{N}}}
	\prod_k d_k\ket{0_c},
\end{equation}
where $\mathcal{N}$ is a normalization constant. However, this construction is not always well defined: in particular, when the Bogoliubov matrix $V$ is singular, some quasiparticle operators annihilate the fermionic vacuum identically, causing the above product to vanish. This situation frequently arises in systems with Majorana zero modes and therefore requires a more careful treatment of the quasiparticle structure.

To resolve this issue, we employ the Bloch--Messiah decomposition, which provides a canonical factorization of the Bogoliubov transformation. Using singular value decomposition, the matrices $U$ and $V$ can be written as
\begin{equation}
	U = C\,\bar{U}\,D^\dagger,
	\qquad
	V = C^*\,\bar{V}\,D^\dagger,
\end{equation}
where $C$ and $D$ are unitary matrices. The matrices $\bar{U}$ and $\bar{V}$ take a block-diagonal canonical form,
\begin{equation}
	\bar{U} =
	\begin{pmatrix}
		1 & & \\
		& \oplus_k u_k \sigma_0 & \\
		& & 0
	\end{pmatrix},
	\qquad
	\bar{V} =
	\begin{pmatrix}
		0 & & \\
		& \oplus_k v_k \sigma_x & \\
		& & 1
	\end{pmatrix},
\end{equation}
with $u_k^2 + v_k^2 = 1$. This decomposition separates the quasiparticle modes into three distinct classes: empty modes, paired modes forming $2\times2$ blocks, and fully occupied modes, a classification that makes it possible to construct the many-body ground state in a manner that remains well defined even when $V$ is singular.

With this structure, the many-body ground state can be written as
\begin{equation}
	\ket{0_d}
	=
	\frac{1}{\sqrt{\mathcal{N}}}
	\prod_{k \in P} \bar{d}_k \bar{d}_{\bar{k}}
	\prod_{k \in O} \bar{d}_k
	\ket{0_c},
\end{equation}
where $P$ denotes the set of paired modes and $O$ denotes the set of fully occupied modes; empty modes are excluded from the product. This construction ensures that all quasiparticle annihilation operators annihilate the ground state, providing a consistent definition of $\ket{0_d}$.

To study dynamical processes, it is often necessary to consider excited states in addition to the ground state. A one-quasiparticle excited state can be defined as
\begin{equation}
	\ket{1_{\bar{d}}}
	=
	\frac{1}{\sqrt{\mathcal{N}}}
	d_N^\dagger
	\prod_{k \notin E} \bar{d}_k
	\ket{0_c},
\end{equation}
where $E$ labels the excited mode. These expressions outline the general procedure for constructing the states
of interest. In our simulations, we use this formalism to construct the
initial state at $t=0$,
\begin{equation}
	\begin{pmatrix}
		c \\
		c^\dagger
	\end{pmatrix}
	=
	\begin{pmatrix}
		U(0) & V^*(0) \\
		V(0) & U^*(0)
	\end{pmatrix}
	\begin{pmatrix}
		d(0) \\
		d^\dagger(0)
	\end{pmatrix},
\end{equation}
and subsequently evolve the Bogoliubov coefficients according to the
time-dependent BdG equation,
\begin{equation}
    i \hbar \frac{\partial}{\partial t} \begin{pmatrix}
        U(t)\\V(t)
    \end{pmatrix} = H_{BdG}(t) \begin{pmatrix}
        U(t)\\V(t)
    \end{pmatrix}.
\end{equation}
The resulting $U(t)$ and $V(t)$ define the instantaneous Bogoliubov
transformation at time $t$. Inverting this transformation expresses the
time-evolved quasiparticle operators in terms of the fixed fermionic basis,
\begin{equation}
\begin{pmatrix}
		d(t) \\
		d^\dagger(t)
	\end{pmatrix}
	=
	\begin{pmatrix}
		U(t) & V^*(t) \\
		V(t) & U^*(t)
	\end{pmatrix}^\dagger
	\begin{pmatrix}
		c \\
		c^\dagger
	\end{pmatrix},
\end{equation}
which is in turn used to construct the time-evolved state $\ket{\psi(t)}$
via the definitions above. Using these definitions, the overlap between excited states at different times can be written as
\begin{widetext}
\begin{align}
	\bra{1_{\bar{d}}(0)}\ket{1_{\bar{d}}(t)}
	&=
	\frac{1}{\sqrt{\mathcal{N}(0)\mathcal{N}(t)}}
	\bra{0_c}
	\left(
	d_N^\dagger(0)
	\prod_{k \notin E} \bar{d}_k(0)
	\right)^\dagger
	d_N^\dagger(t)
	\prod_{k \notin E} \bar{d}_k(t)
	\ket{0_c} \nonumber \\
	&=
	\frac{(-1)^{n_d(n_d-1)/2}}{\sqrt{\mathcal{N}(0)\mathcal{N}(t)}}
	\bra{0_c}
	\left(
	\prod_{k \notin E} \bar{d}^\dagger_k(0)
	\right)
	d_N(0) d_N^\dagger(t)
	\left(
	\prod_{k \notin E} \bar{d}_k(t)
	\right)
	\ket{0_c},
\end{align}
\end{widetext}
where $n_d$ denotes the total number of quasiparticle operators involved, and the overall sign arises from the fermionic anticommutation relations.

The evaluation of such expressions is carried out using the generalized Wick's theorem, which states that expectation values of products of fermionic operators can be expressed as a Pfaffian. Accordingly, the overlap takes the compact form
\begin{equation}
	\bra{1_{\bar{d}}(0)}\ket{1_{\bar{d}}(t)}
	=
	\frac{(-1)^{n_d(n_d-1)/2}}{\sqrt{\mathcal{N}(0)\mathcal{N}(t)}}
	\, \mathrm{pf}\!\left[A(t)\right],
\end{equation}
where $A(t)$ is an antisymmetric matrix whose elements are given by all possible pairwise contractions between the quasiparticle operators at times $0$ and $t$. The matrix $A(t)$ is constructed from the relations between the quasiparticle operators and the underlying fermionic operators encoded in the Bogoliubov transformation.

This Pfaffian-based formulation provides a numerically efficient and robust method for computing many-body overlaps in time-dependent BdG systems\cite{wimmer_algorithm_2012}. It is particularly well suited for studying Majorana bound states, where zero-energy modes and parity effects render simpler overlap constructions ill defined, and it is this formalism that we use throughout to compute ground-state fidelity, excited-state overlaps, and dynamical error measures under various driving protocols.

\bibliography{braid.bib}
\end{document}